\documentstyle[amssymb,12pt,epsfig]{article}
\begin{document}
\baselineskip=22pt

~~~~~~~~~~~~~~~~~~~~~~~~~~~~~~~~~~~~~~~~~~~~~~~~~~~~~~~~~~~~~~~~~~~~~~~~~~~~~~~~~~~~~~~~~~~~BIHEP-TH-2003-38

\vspace{0.6cm}

\begin{center}
{\Large \bf Noncommutative Theory in Light of Neutrino Oscillation
}

\bigskip

Shao-Xia Chen\footnote{ruxanna@mail.ihep.ac.cn}~~and~~Zhao-Yu Yang\footnote{yangzhy@mail.ihep.ac.cn}\\
{\em Institute of High Energy Physics,
Chinese Academy of Sciences} \\
{\em P.O.Box 918(4), 100039 Beijing, China}
\end{center}

\vspace{1.2cm}
\begin{abstract}
Solar neutrino problem and atmospheric neutrino anomaly which are
both long-standing issues studied intensively by physicists in the
past several decades, are reckoned to be able to be solved
simultaneously in the framework of the assumption of the neutrino
oscillation. For the presence of the Lorentz invariance in the
Standard Model, the massless neutrino can't have flavor mixing and
oscillation. However, we exploit the $q$-deformed noncommutative
theory to derive a general modified dispersion relation, which
implies some violation of the Lorentz invariance. Then it is found
that the application of the $q$-deformed dispersion relation to
the neutrino oscillation can provide a sound explanation for the
current data from the reactor and long baseline experiments.
\end{abstract}

\vspace{0.6cm}

\vspace{1.2cm}

The current knowledge concerning high energy neutrinos mainly
comes from solar, atmospheric and long baseline neutrino
experiments, which all present some inconsistence with the
electro-weak Standard Model. In the Super-Kamiokande (S-K)
experiment \cite{sk, sk2}, the results of directly detecting
$\nu_ee$ scattering via the observation of the Cherenkov light in
the large water Cherenkov detector, and the detection in the SNO
\cite{SNO1}-\cite{SNO3} using a heavy water Cherenkov detector
have significantly confirmed that the number of solar neutrinos
detected in these experiments is less than that predicted by the
theories of the standard solar model. Some experiments detecting
atmospheric neutrino, including several water Cerenkov
experiments(Kamiokande \cite{Kamiokande}, S-K and IMB \cite{IMB})
together with the iron calorimeter Soudan 2 experiment
\cite{Soudan}, observing atmospheric neutrinos over energies
varying from sub-GeV to tens of GeV have shown obvious discrepancy
between the expected and measured ratio of the numbers of muon and
electron events $R_{\mu/e}$. As one of the new generation of long
baseline neutrino oscillation searches, KamLAND \cite{KamLAND} has
published its first results recently, which give strong evidence
for the disappearance of neutrinos travelling from a power reactor
to a far detector. Meanwhile, the results of long baseline K2K
experiment \cite{k2k} imply a reduction of $\nu_{\mu}$ flux
together with a distortion of the energy spectrum. Thanks to all
the experiments above, it is now convincing that the long-standing
solar neutrino $\nu_e$ deficit and the atmospheric neutrino
$\nu_{\mu}$ anomaly are both from neutrino oscillations between
the flavor eigenstates.

Neutrino oscillations may arise, in general, when some terms are
added to the neutrino sector of the Standard Model Lagrangian in
the way generating the discordance of the eigenstates of the total
Hamiltonian with the neutrino flavor eigenstates. In addition to
the general scenario of neutrino oscillations
\cite{mass1}-\cite{mass5} from the mismatch between the
diagonalization of the charged lepton mass matrix and that of the
neutrino mass matrix in an arbitrary flavor basis, there have
appeared considerable papers \cite{LIV1}-\cite{LIV4} discussing
the neutrino oscillation in the context of the premise that the
oscillations may be the effects of the disagreement of the flavor
eigenstates with the velocity eigenstates which are defined as the
energy eigenstates of the massless neutrino and much has already
been achieved. In the perspective of the violation of the Lorentz
invariance several kinds of outcomes have appeared in the
literature: $\lambda\propto E^0$ \cite{glashow1-0, 0},
$\lambda\propto E^{-1}$ \cite{1, 12}, $\lambda\propto E^{-2}$
\cite{LIV1, 2, 22} and $\lambda\propto E^{-3}$ \cite{3}, here
$\lambda$ is the neutrino oscillation length which will be defined
in the subsequent discussion and $E$ is the mean value of observed
neutrino energy.

In this Letter, we will start from a modified dispersion relation
induced by the $q$-deformed noncommutative theory which also
indicates the violation of the Lorentz invariance, to investigate
the oscillations between the different flavor eigenstates of the
massless neutrinos.

The $q$-deformed noncommutative theory, as one of the significant
application of the quantum group \cite{quantum1}-\cite{qg5} to the
modern physics, plays a useful role in the study of the high
energy physical phenomena. In the $q$-deformed noncommutative
theory physical quantity $[x]$ is defined in terms of a parameter
$q$ (taken to be real for simplicity):
$$
[x]=\displaystyle\frac{q^x-q^{-x}}{q-q^{-1}}~.
$$
The parameter $q$ measures the degree of the deviation of the
considered system from the usual commutative case. Now that
neutrino oscillation is a physical phenomenon related to the GeV
scale energy, it is natural to make use of the $q$-deformed
noncommutative theory to investigate it. The dispersion relation
for a fermion in the $q$-deformed case is of the form \cite{chen1,
chen2}:
\begin{equation}
\label{dispersion}
E=\sqrt{m^2+p^2}+\frac{\sqrt{m^2+p^2}(4m^2+4p^2-\omega^2)}
   {24\omega^2}(q-1)^2.
\end{equation}
For the purpose of subsequent discussion, we derive the
approximated form of Eq. (\ref{dispersion}) for massless
neutrinos:
\begin{equation}
\label{approximation}
E=p\left(1+\frac{(q-1)^2p^2}{6{\omega}^2}\right)=p(1+\alpha p^2)~,
\end{equation}
here the notation
$\alpha$=$\displaystyle\frac{(q-1)^2}{6\omega^2}$ has been
introduced and characterizes the Lorentz invariance violation in
$q$-deformed noncommutative theory.

As argued in \cite{glashow1-0, glashow2}, the Lorentz invariance
violation would indicate that neutrinos may differ in their
maximal attainable velocities, which disagree with the light
velocity $c$ in vacua. In this senario, neutrino oscillations can
occur if the neutrino flavor eigenstates decided by the weak
interactions are coherent superposition of the neutrino velocity
eigenstates. In the case of two neutrino flavors, the flavor
eigenstates can be expressed in terms of the velocity eigenstates:
\begin{equation}
\label{mixing}
\nu_{\mu}=\nu_1\cos{\theta}+\nu_2\sin{\theta}, ~~~\\
\nu_e=\nu_2\cos{\theta}-\nu_1\sin{\theta}~.
\end{equation}
here $\theta$ is the mixing angle between different flavor
eigenstates. The probability that neutrino oscillates from
$\alpha$ flavor to $\beta$ flavor is:
\begin{equation}
\label{probability}
P_{\alpha\rightarrow\beta}=\sin^22\theta\sin^2\left(\frac{{\pi}L}{\lambda}\right)~,
\end{equation}
where $L$ is the cosmological distance travelled by neutrinos
between the emission and the detection, and the oscillation length
$\lambda$ is defined as
$\lambda=\displaystyle\frac{2\pi}{E_1-E_2}$. The mixing angle
$\theta$ only determines the amplitude of the oscillation but not
affect the oscillation length. It should be noted that, although
here we consider merely the simplified case of two neutrino
flavors, however, in the case of more than two flavors, Eq.
(\ref{probability}) will take a more complicated form but not
contain more intrinsical physical contents, so we will concentrate
only on the two flavors case.

During their evolution, neutrinos propagate as a linear
superposition of their velocity eigenstates whose energy
eigenvalues are $E_a$ ($a=1,~2$). Exploiting Eq.
(\ref{approximation}), one obtains that the energy difference is
\begin{equation}
\label{energy} E_1-E_2=\delta{\alpha}E^3~,
\end{equation}
where $\delta{\alpha}$=$\displaystyle\alpha_1-\alpha_2$. Combining
the definition of the oscillation length $\lambda$, (\ref{energy})
and (\ref{probability}), one can acquire
\begin{equation}
\label{q-length} \lambda=\frac{2\pi}{\delta\alpha
E^3}~,~~~~~~~P_{\alpha\rightarrow\beta}=\sin^22\theta\sin^2\left(\frac{LE^3\delta{\alpha}}{2}\right)~.
\end{equation}
It is obvious that the oscillation length $\lambda$ is of the form
$\lambda\propto E^{-3}$, which is identical with that appeared in
the literature \cite{3}. For a ulterior discussion, we can plot
several curves in {\bf Fig.1} describing the dependence of the
neutrino oscillation length $\lambda$ on $\delta\alpha$ for
different energies. From the curves we find that only a tiny
deviation of $q$ from 1 can unravel the existent data, which means
that in the phenomenon of neutrino oscillation, at least on the
current energy scale of detection, the degree of noncommutativity
is very small. Of course, we must note that here we only take into
account the neutrino oscillation from the disagreement of the
neutrino flavor eigenstates with velocity eigenstates and omit the
conventional oscillation mechanism in which neutrino oscillation
comes from the difference between neutrino mass eigenstates and
flavor eigenstates. However, it is evident that in the presence of
neutrino mass the requisite deviation of $q$ from 1 will be
smaller than that for massless neutrinos. In general, if neutrinos
have mass, they may have simultaneous velocity and mass
oscillations. Then exploiting the argument in \cite{glashow1-0,
glashow2}, one can further discuss the situation in the presence
of two oscillations.

\begin{center}
{\bf Experimental Data}\\
~\\
\begin{tabular}{|c|c|c|c|}
\hline EXP. & STATUS & $\left<E\right>$(GeV) & L(km) \\
\hline CHORUS & closed 1997 & 26 & 0.85  \\
\hline NOMAD & closed 1999 & 24 & 0.94 \\
\hline SK & operating & 1.3 & $10-10^4$ \\
\hline K2K & operating & 1.3 & 250 \\
\hline SNO & operatng & 0.008 & $10^8$\\
\hline MINOS & starting 2003 & 15 & 730\\
\hline CNGS & starting 2005 & 17 & 732\\
\hline
\end{tabular}
\end{center}
{\bf Table 1}. {\small Shown for each experiment are its operation
status, mean value of observed neutrino energy and typical
neutrino flight distance $L$. The table is quoted from the paper
\cite{LIV1}.}

\begin{figure}[hbtp]
\begin{center}
\epsfig{file=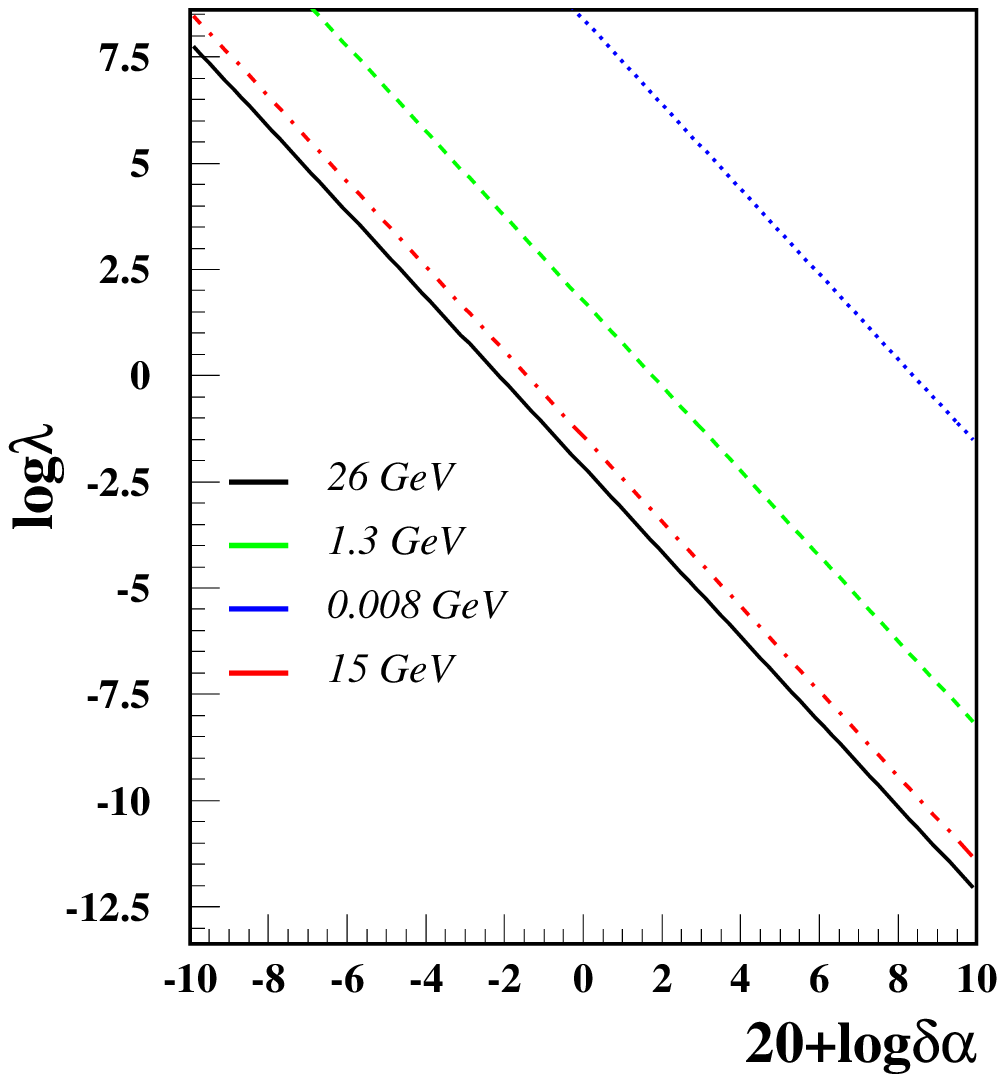,height=12.cm,angle=0.} \end{center}
\begin{center}{\bf Fig 1. }Oscillation length for different
energies
\end{center}
\end{figure}

To summarize, we exhibit in this Letter the relation between the
neutrino oscillation length $\lambda$ and the Lorentz invariance
violation parameter $\delta\alpha$, which can be verified by the
existent and future neutrino oscillation experiments. Especially,
when the data from long baseline experiments and neutrino
factories come to be available, Lorentz invariance violation in
neutrino oscillation can be probed to new and significant levels.
Once the oscillation length are measured, $\delta\alpha$ will be
derived as additional contribution apart from the observations of
ultrahigh energy cosmic rays \cite{chen2} to determine the
$q$-deformed noncommutativity parameter.

\centerline{\large\bf Acknowledgements} The work was supported in
part by the Natural Science Foundation of China.

\end{document}